\documentclass{article}
\usepackage{authblk}
\usepackage{hyperref}
\usepackage{multirow}
\usepackage{array}
\begin{document}
\date{}
\title{Powering the Internet of Things with RIOT: Why? How? What is RIOT?}
\author{E. Baccelli\footnote{Emmanuel Baccelli works for INRIA. He is hosted at FU Berlin for joint work on RIOT.} \hspace{0.08cm} and K. Schleiser }
\maketitle

The crucial importance of software platforms was highlighted by recent events both at the political level (e.g. renewed calls for digital data and operating system ''sovereignty'', following E. Snowden's revelations) and at the business level (e.g. Android generated a new industry worth tens of billions of euros yearly). In the IoT, which is expected to generate business at very large scale, while threatening even more individual privacy, such aspects will be exacerbated. The need for an operating system like RIOT \cite{RIOT-website} \cite{RIOT-ercim} stems from this context, as outlined in the following.

\section{RIOT: Why?}
\label{why}
Despite enormous expectations for the Internet of Things \cite{Cisco}, there is so far no de facto standard software platform to program most IoT devices, which are constrained in resources such as CPU, memory, energy. IoT software platforms thus face conflicting requirements : interoperability with the Internet, memory-constrained embedded programming, and portable, open-source code. Such limitations and requirements are expected to last \cite{Moore}, and make impossible the use of traditional platforms such as Linux or derivatives, e.g. Android. In effect, the anticipated IoT industry depends on the emergence of standard software platforms for IoT devices. In this context, it is vital that a strong alternative emerges which can achieve the following goals:

\begin{itemize}
\item allow long-term IoT software security and robustness, 
\item enable trust, transparency and the protection of IoT users' privacy,
\item accelerate innovation by spreading IoT software development costs,
\item reduce environmental impact by preventing IoT device lock-down. 
\end{itemize}

\newpage
\noindent Experience over the last decades (for instance, with Linux) shows that such goals are most likely achieved with a software platform that is: 

\begin{itemize}
\item open source, 
\item free,
\item driven by a grassroots community.
\end{itemize}

\section{RIOT: How?}

In order to achieve the goals listed above in Section \ref{why}, the RIOT community gathers a large number of contributors \cite{RIOT-contributors} from around the world (with various backgrounds including industry, academia, and hobbyists) and uses an approach based on the following building blocks and principles.
\ \\ \ \\
The RIOT community self-organizes in oder to follow clearly defined open processes \cite{RIOT-processes}, which favor transparency, and code quality. Technically, the RIOT community uses an open, online tool (GitHub \cite{RIOT-GitHub}) allowing large-scale distributed revision control and source code management. Furthermore, RIOT code contribution processes mandate that the core of the software platform code is free and licensed with a non-viral copyleft license (LGPLv2.1), which is expected to avoid as much as possible death-by-forking while allowing indirect business models around RIOT (similar to business around Linux). 


\section{What is RIOT?}

RIOT is an open source operating system which achieves the combination of (i) the necessary memory and energy efficiency to fit on the widest range of relevant low-end IoT devices \cite{IoT-OS-Survey}, while (ii) offering the functionalities of a full-fledged operating system, i.e. full-featured, extensible network stacks providing spontaneous wireless networking and end-to-end Internet connectivity (e.g. from IoT device to the cloud), as well a powerful API providing state-of-the-art development and debugging tools, uniform and consistent across all supported hardware spanning across 32-bit, 16-bit, and 8-bit architectures. For an technical overview of competing IoT software platforms, one can refer to this survey~\cite{IoT-OS-Survey}.
\ \\ \ \\
From a technical perspective, in a nutshell, RIOT is an operating system based on a micro-kernel architecture \cite{RIOT-infocom}, with built-in energy efficiency and real-time capabilities. RIOT enables full multi-threading as usually experienced on traditional operating systems (e.g. Linux) and offers the capability to develop efficient IoT applications in standard  programming languages (C and C++ ), with tools that are well-known such as gcc, gdb, valgrind, wireshark etc. Instance(s) of RIOT can actually run as process(es) on Linux and Mac OS, and this feature is heavily used for debugging and testing purposes. Such characteristics eliminate most learning curves for embedded software design and shorten development life-cycles for IoT products. For a third-party code analysis of the RIOT code base, one can refer to \cite{RIOT-black-duck}.
\ \\ \ \\
On top of the aforementioned features, RIOT supports connectivity via several network stacks. One stack (GNRC) offers low-power IPv6 and 6LoWPAN standard-compliant connectivity, with an extendable software architecture. In particular, the modularity of GNRC enables efficient plug-in of newer and upcoming IoT protocols which go beyond the minimal compliant standard (e.g. RPL extensions \cite{P2P-RPL-RFC} \cite{P2P-RPL-paper}). Other stacks are available, including for instance an experimental stack based on novel network paradigms currently studied by the research community, such as content-centric networking \cite{ICN-IoT-paper}.
\ \\ \ \\
Due to the characteristics described in this document, RIOT offers an attractive base for embedded IoT software development, may it be for industrial purposes, for prototyping, for experimental research purposes, or for teaching in this domain -- similarly to Linux in another domain.

\end{document}